\documentclass[aps,prl,twocolumn]{revtex4}%
\usepackage{amsfonts}
\usepackage{amsmath}
\usepackage{amssymb}
\usepackage{graphicx}%
\providecommand{\U}[1]{\protect\rule{.1in}{.1in}}

\begin{document}
\title{Alternative approach to computing transport coefficients: application to
conductivity and Hall coefficient of hydrogenated amorphous silicon}
\author{Ming-Liang Zhang and D. A. Drabold}
\affiliation{Department of Physics and Astronomy, Ohio University, Athens, Ohio 45701}

\begin{abstract}
We introduce a theoretical framework for computing transport coefficients for
complex materials. As a first example, we resolve long-standing
inconsistencies between experiment and theory pertaining to the conductivity
and Hall mobility for amorphous silicon and show that the Hall sign anomaly is
a consequence of localized states. Next, we compute the AC conductivity of
amorphous polyanaline. The formalism is applicable to complex materials
involving defects and band-tail states originating from static topological
disorder and extended states. The method may be readily integrated with
current \textit{ab initio} methods.

\end{abstract}
\date{March 30, 2010}
\maketitle


The atomistic understanding of the electrical conductivity\cite{ma,kik} and
Hall coefficient\cite{fri71,bot77,Emin77,fri78,mov} are key unsolved problem
in the physics of amorphous semiconductors\cite{str}. The challenge is due to
two factors: (1) a complex array of localized states of varying physical
origin; and (2) at moderate temperature, both localized states and extended
states are accessible excited states\cite{motda,epjb}.

Previous work on the conductivity\cite{hII,lan62} and Hall
coefficient\cite{h61,fri63,fir,sch65,hol68} of amorphous semiconductors has
involved either Miller-Abrahams theory or small polaron models\cite{motda}.
Neither approach is ideal for amorphous semiconductors, with their complex
menagerie of localized states due to defects, and tail states due to
topological\cite{urbachletter} or chemical disorder\cite{str} and
electron-phonon couplings strongly dependent upon carrier
localization\cite{attafynn}.

In this Letter, we first develop a novel formalism for the linear response to
a mechanical perturbation\cite{ku57}. The method properly includes the four
possible transitions between extended or localized initial and final states.
The full results and many intermediate details are provided in Ref.\cite{4c};
here we require only transitions between localized states. Our work resolves
the puzzle of the sign anomaly of the Hall mobility in a-Si:H:, provides the
temperature dependence of the transport coefficients, and accurately predicts
the ac conductivity of polyaniline. The new formalism is expected to be
applicable well beyond the examples of this Letter and lends itself to
inclusion in current \textit{ab initio} schemes.

In the small polaron hopping regime, the Kubo linear response
formula\cite{ku57} has been used to compute conductivity and Hall
mobility\cite{lan62,fir,sch65,hol68}. The key mathematical obstacle to
computing the current-current correlation function is the imaginary time
integral,\cite{lan62,fir,sch65,hol68} which results from the commutator
between the microscopic current and density matrix\cite{ku57}. For a
``mechanical perturbation" (for which an external disturbance may be expressed
with additional terms in the Hamiltonian\cite{ku57}), we can avoid this
troublesome commutator, and within this picture the local density operator
$\widehat{\rho}$ of a quantity (charge, energy etc.) can be easily
constructed\cite{irv}.

We first average $\widehat{\rho}$ over a state $\Psi^{\prime}(t)$ of the
system with the mechanical perturbation, from which the \textit{microscopic}
local density $\rho(\mathbf{r},t)=\langle\Psi^{\prime}(t)|\widehat{\rho}%
|\Psi^{\prime}(t)\rangle$ is obtained. Next we calculate $\partial
\rho(\mathbf{r},t)/\partial t$ by means of the time-dependent Schrodinger
equation $i\hbar\partial\Psi^{\prime}(t)/\partial t=H^{\prime}(t)\Psi^{\prime
}(t)$, where $H^{\prime}(t)$ is the total Hamiltonian of [system + mechanical
perturbation]. The third step is to apply the local density (charge density,
energy density) continuity equation: $\partial\rho(\mathbf{r},t)/\partial
t+\nabla\cdot$ $\mathbf{j}_{m}(\mathbf{r},t)=0$; the \textit{microscopic}
response $\mathbf{j}_{m}$ (current density, energy flux etc.) is then
expressed in terms of $\Psi^{\prime}(t)$. Practically speaking, $\Psi^{\prime
}(t)$ may be computed to the required order with perturbation theory. By
substituting $\Psi^{\prime}(t)$ into the expression for $\mathbf{j}%
_{m}(\mathbf{r},t)$, one can obtain the \textit{microscopic} response to the
required order of mechanical disturbance. Spatial and ensemble average are
taken at the final stage. The desired transport coefficients can then be
extracted from the ensemble average of the spatially averaged flux
$\mathbf{j}$.

Since the state of the system is completely determined from the initial
conditions, averaging over initial state can be delayed until the final stage.
Thus we can avoid the commutator between flux and density matrix,
\textit{i.e.} the integral over imaginary time.

Consider then,  a system with $N_{e}$ electrons and $\mathcal{N}$ nuclei in the
presence of an electromagnetic field with potentials ($\mathbf{A},\phi$), the
charge density of state $\Psi^{\prime}$ at $\mathbf{r}$ $\rho^{\prime
}(\mathbf{r},t)=\int d\tau\Psi^{\prime\ast}\widehat{\rho}(\mathbf{r}%
)\Psi^{\prime}$, where the arguments of wave function $\Psi^{\prime}$ are
$(\mathbf{r}_{1}\cdots\mathbf{r}_{N_{e}};\mathbf{W}_{1}\cdots\mathbf{W}%
_{\mathcal{N}};t)$, $\mathbf{W}_{1}$ is the position of the first nucleus,
etc. $\widehat{\rho}(\mathbf{r})=\sum_{j}e\delta(\mathbf{r}-\mathbf{r}%
_{j})-\sum_{L}Z_{L}e\delta(\mathbf{r}-\mathbf{W}_{L})$ is the charge density
operator. $d\tau=d\mathbf{r}_{1}\cdots d\mathbf{r}_{N_{e}}d\mathbf{W}%
_{1}\cdots d\mathbf{W}_{\mathcal{N}}$ is the volume element in configuration
space. The evolution of the state is determined by the time-dependent
Schrodinger equation for which the total Hamiltonian includes the interaction
between system and external field. The contribution from the electrons is:%
\[
\mathbf{j}_{m}(\mathbf{r},t)=\frac{i\hbar eN_{e}}{2m}\int d\tau^{\prime}%
(\Psi^{\prime}\nabla_{\mathbf{r}}\Psi^{\prime\ast}-\Psi^{\prime\ast}%
\nabla_{\mathbf{r}}\Psi^{\prime})
\]%
\begin{equation}
-\frac{e^{2}N_{e}}{m}\mathbf{A}(\mathbf{r},t)\int d\tau^{\prime}\Psi
^{\prime\ast}\Psi^{\prime}, \label{mc}%
\end{equation}
where the arguments of $\Psi^{\prime}$ are $(\mathbf{r,r}_{2},\cdots
,\mathbf{r}_{N_{e}};\mathbf{W}_{1},\mathbf{W}_{2},\cdots,\mathbf{W}%
_{\mathcal{N}};t)$, $d\tau^{\prime}=d\mathbf{r}_{2}\cdots d\mathbf{r}_{N_{e}%
}d\mathbf{W}_{1}d\mathbf{W}_{2}\cdots d\mathbf{W}_{N_{n}}$. In Eq.(\ref{mc}),
the anti-symmetry of $\Psi^{\prime}$ under exchange of particles was used, and
the Coulomb gauge $\nabla\cdot\mathbf{A}(\mathbf{r},t)=0$ was adopted to
simplify the expression. $\mathbf{j}_{m}(\mathbf{r},t)$ is gauge
invariant\cite{bcs,fri63}. Without the nuclear coordinates, Eq.(\ref{mc}) is
identical to the form used by BCS to compute paramagnetic and diamagnetic
currents for superconductors in which Bloch states are not radically modified
by the electron-phonon (e-ph) interaction\cite{bcs}. Eq.(\ref{mc}) is a
generalization to arbitrary strength of e-ph interaction, and may be used for
the localized carriers in amorphous semiconductors or the polarons in ionic
and molecular crystals.

We now apply Eq.(\ref{mc}) to compute the conductivity and Hall mobility of an
amorphous semiconductor. The ratio of the second term to the first term is
$eA/p$ ($p$ is electron momentum), so that the contribution from the 2nd term
may be neglected. Since the carrier concentration is low in lightly-doped
amorphous semiconductors, one can invoke the single-electron approximation,
and $\Psi^{\prime}(t)$ for $\mathbf{j}_{m}$ may be replaced by the
single-electron wave function $\psi^{\prime}(\mathbf{r};x_{1},x_{2}%
,\cdots,x_{3\mathcal{N}};t)$, where $\mathbf{r}$ is the coordinate of the
carrier, $x_{1},x_{2},\cdots,x_{3\mathcal{N}}$ are the displacements of
$3\mathcal{N}$ vibrational degrees of freedom.

Using perturbation theory, one can expand $\psi^{\prime}(t)$ to the required
order of external field\cite{epjb,4c}. We use A with subscripts to label
localized states, denote the coupling between two localized states $\phi_{A}$
and $\phi_{A_{1}}$ caused by external field as $J_{A_{1}A}^{field}=\int
d\mathbf{r}\phi_{A_{1}}^{\ast}h_{fm}\phi_{A}$, where $h_{fm}=(i\hbar
e/m)\mathbf{A}(\mathbf{r})\cdot\nabla_{\mathbf{r}}+e^{2}\mathbf{A}%
^{2}(\mathbf{r})/(2m)+e\phi(\mathbf{r})$. The spatially averaged
\textit{microscopic} current density to second order of $J^{field}$ is%
\[
\mathbf{j}(\mathbf{s},t)=-\frac{N_{e}\hbar e}{m\Omega_{\mathbf{s}}}%
\int_{\Omega_{\mathbf{s}}}d\mathbf{r}\int[%
{\displaystyle\prod\limits_{j=1}^{3\mathcal{N}}}
dx_{j}]
\]%
\[
\{\operatorname{Im}(\psi^{(0)}\nabla_{\mathbf{r}}\psi^{(1)\ast}-\psi^{(1)\ast
}\nabla_{\mathbf{r}}\psi^{(0)})
\]%
\begin{equation}
+\operatorname{Im}(\psi^{(0)}\nabla\psi^{(2)\ast}-\psi^{(2)\ast}\nabla
\psi^{(0)}+\psi^{(1)}\nabla\psi^{(1)\ast})\}, \label{1e1}%
\end{equation}
where $\psi^{(1)}$ is change in state to order $J^{field}$, and $\psi^{(2)}$
is change in state to order $[J^{field}]^{2}$, where $\psi^{(0)}(t)$ is the
state of carrier at time $t$ \textit{without} external field. $\Omega
_{\mathbf{s}}$ is the \textquotedblleft physical infinitesimal" volume of
Kubo\cite{kubo}. Because the initial state of the phonon-dressed carrier is
unknown, we need to average $j_{k}(k=x,y,z)$ over initial phonon distribution
and single-electron states.

To compute the conductivity, we only require the order $J^{field}$ term of
Eq.(\ref{1e1}). If one applies a voltage drop across the material, the
potentials are $\mathbf{A}=0$ and $\phi=-2\mathbf{E}_{0}\cdot\mathbf{r}%
\cos\omega t$. Using perturbation theory, one can compute $\psi^{(0)}$ and
$\psi^{(1)}$ to order $J^{1}$. Substitute $\psi^{(0)}$ and $\psi^{(1)}$ into
the first term of Eq.(\ref{1e1}), the conductivity from the LL transition
is\cite{4c}:
\[
\left.
\begin{array}
[c]{c}%
\operatorname{Re}\sigma_{\alpha\beta}(\omega)\\
\operatorname{Im}\sigma_{\alpha\beta}(\omega)
\end{array}
\right\}  =\frac{N_{e}e^{2}}{\Omega_{\mathbf{s}}}\operatorname{Im}%
i\sum_{AA_{1}}[I_{A_{1}A+}\pm I_{A_{1}A-}]
\]%
\[
\times f(E_{A}^{0})[1-f(E_{A_{1}}^{0})]v_{A_{1}A}^{k\ast}(E_{A}^{0}-E_{A_{1}%
}^{0})^{-1}(w_{AA_{1}}^{j}-v_{A_{1}A}^{j}),
\]%
\[
+\frac{N_{e}e^{2}}{\Omega_{\mathbf{s}}}\operatorname{Im}\sum_{AA_{1}A_{3}%
}f(E_{A}^{0})[1-f(E_{A_{1}}^{0})][1-f(E_{A_{3}}^{0})]
\]%
\[
\times(E_{A}^{0}-E_{A_{1}}^{0})^{-1}\hbar^{-1}J_{A_{3}A}[I_{A_{3}A_{1}A+}\pm
I_{A_{3}A_{1}A-}]
\]%
\begin{equation}
\times(w_{A_{3}A_{1}}^{j}-v_{A_{1}A_{3}}^{j})v_{A_{1}A}^{k\ast}\label{scd}%
\end{equation}%
\[
+\frac{N_{e}e^{2}}{2\hbar\Omega_{s}}\sum_{A_{2}A_{1}A}\operatorname{Im}%
(w_{AA_{2}}^{\beta}-v_{A_{2}A}^{\beta})(E_{A_{1}}^{0}-E_{A_{2}}^{0}%
)^{-1}(v_{A_{2}A_{1}}^{\alpha})^{\ast}J_{A_{1}A}^{\ast}%
\]%
\[
(Q_{1A_{2}A_{1}A+}\pm Q_{1A_{2}A_{1}A-})f(E_{A}^{0})[1-f(E_{A_{2}}^{0})]
\]%
\[
+\frac{N_{e}e^{2}}{2\hbar\Omega_{\mathbf{s}}}\sum_{A_{2}A_{1}A}%
\operatorname{Im}(w_{AA_{2}}^{\beta}-v_{A_{2}A}^{\beta})J_{A_{2}A_{1}}^{\ast
}(E_{A}^{0}-E_{A_{1}}^{0})^{-1}(v_{A_{1}A}^{\alpha})^{\ast}%
\]%
\[
(Q_{2A_{2}A_{1}A+}\pm Q_{2A_{2}A_{1}A-})f(E_{A}^{0})[1-f(E_{A_{2}}^{0})]
\]
$\operatorname{Re}\sigma_{jk}(\omega)$ ($j,k=x,y,z$) takes the positive sign
and $\operatorname{Im}\sigma_{jk}(\omega)$ takes the negative sign. Unlike
previous theories\cite{motda},the dc conductivity may be directly extracted
from Eq.(\ref{scd}) without a limiting process. Here, $v_{A_{1}A}^{k}=\int
d\mathbf{r}\phi_{A_{1}}^{\ast}(p_{k}/m)\phi_{A}$ and $w_{AA_{1}}^{j}=\int
d\mathbf{r}\phi_{A}(p_{j}/m)\phi_{A_{1}}^{\ast}$ are the velocity matrix
elements of a carrier. $f(E_{A}^{0})$ is the Fermi distribution. $J_{A_{1}%
A}=\int d\mathbf{r}\phi_{A_{1}}^{\ast}\sum_{p\notin D_{A}}U(r-\mathcal{R}%
_{p})\phi_{A}$ is transfer integral from $\phi_{A}$ to $\phi_{A_{1}}$. In
Eq.(\ref{scd}), the first term is order $J^{0}$ contribution, the second term
is order $J^{1}$ contribution. Here, $I_{A_{1}A\pm}(\omega,T)$ has dimension
of time (denote its order as $t_{A_{1}A}^{\pm}$), and reflects the time evolution
of $\psi^{\prime}(t)$ in a field-driven 2-site transition, cf. Eq.(C1) in
\cite{4c}. The characteristic time $t_{A_{1}A}^{\pm}$ may be interpreted as
the mean free carrier time. Similarly the two-fold time integrals
$I_{A_{3}A_{1}A\pm}$, $Q_{1A_{2}A_{1}A\pm}$, and $Q_{1A_{2}A_{1}A\pm}$, have
dimension of [time]$^{2}$, reflecting the time evolution of $\psi^{\prime}(t)$ in
a 3-site transition induced by transfer integral $J$; they have same order of
magnitude $[t_{A_{3}A_{1}A}^{\pm}]^{2}$, cf. Eqs.(C3,C7-C9) in \cite{4c}.
$\hbar^{-1}J_{A_{3}A}[t_{A_{3}A_{1}A}^{\pm}]^{2}$ is the mean
free time in a 3-site process. The interference between $J^{field}$ and $J$ is
displayed between two components (reached through
different paths) of the final state\cite{4c}. At high temperature
$k_{B}T>\hbar\overline{\omega}$ ($\overline{\omega}$ is first peak in phonon
spectrum), the time integrals $I_{s}$ and $Q_{s}$ can be approximately
computed using the method of steepest descent (see Appendix D in \cite{4c}).
For example:%
\[
I_{A_{1}A\pm}=e^{-\beta\hbar(\omega_{A_{1}A}^{\prime}\mp\omega)/2-\beta
\lambda_{A_{1}A}/4}%
\]%
\[
\{C^{-1/2}\sum_{n=0}^{\infty}\frac{(-i)^{n}(\omega-\omega_{A_{1}A}^{\prime
})^{n}}{n!C^{n/2}}\Gamma(\frac{n+1}{2})
\]%
\begin{equation}
-i[\sum_{\alpha}\frac{1}{2}(\theta_{\alpha}^{A_{1}}-\theta_{\alpha}^{A}%
)^{2}\omega_{\alpha}\cosh\frac{\beta\hbar\omega_{\alpha}}{2}]^{-1}%
\},\label{0+}%
\end{equation}
where $\theta_{\alpha}^{A_{1}}$ is the shift in origin of the $\alpha^{th}$
normal mode caused by a carrier in state $\phi_{A_{1}}$, $\lambda_{A_{1}%
A}=\frac{1}{2}\sum_{\alpha}\hbar\omega_{\alpha}(\theta_{\alpha}^{A_{1}}%
-\theta_{\alpha}^{A})^{2}$ and $C=\hbar^{-2}k_{B}T\lambda_{A_{1}A}%
$\cite{epjb,4c}. The mean free time decreases with increasing
$\lambda_{A_{1}A}$ and energy difference: $t_{A_{1}A}^{\pm}\thicksim
\hbar(k_{B}T\lambda_{A_{1}A})^{-1/2}e^{-\beta\hbar(\omega_{A_{1}A}^{\prime}%
\mp\omega)/2-\beta\lambda_{A_{1}A}/4}$. The average mobility $\mu$ is defined
by $\sigma_{xx}=e^{2}(N_{e}/\Omega_{\mathbf{s}})\mu$. From Eq.(\ref{scd}), one
can see that $\mu$ depends on the energy distribution and spatial distribution
of localized states. A typical value of $\mu$ can be estimated: $\mu\thicksim
v_{A_{1}A}^{2}t_{A_{1}A}\left(  E_{A}^{0}-E_{A_{1}}^{0}\right)  ^{-1}%
+v_{A_{1}A}v_{A_{3}A_{1}}(t_{A_{3}A_{1}A}^{2}\hbar^{-1}J_{A_{3}A})\left(
E_{A}^{0}-E_{A_{1}}^{0}\right)  ^{-1}$.

\begin{figure}[ptb]
\begin{center}
\includegraphics[width=7cm]{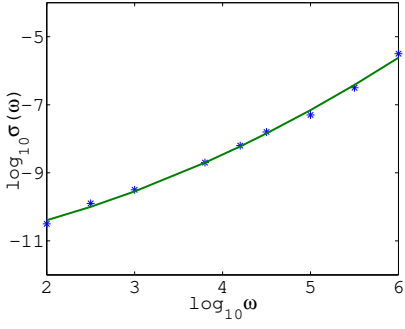}
\end{center}
\caption{AC conductivity of polyaniline as a function of frequency at T=300K:
star symbols denote experimental data\cite{india}, solid line is from the
first three terms in Eq.(\ref{0+}).}%
\label{anline}%
\end{figure}

As a test, we apply Eqs.(\ref{scd},\ref{0+}) to the frequency dependence of
the ac conductivity in polyaniline at T=300K\cite{india}. The Austin-Mott
$\omega^{0.8}$ law\cite{motda} does not accurately fit
experiments\cite{greece}. In Fig.\ref{anline}, we fit the data\cite{india}
with the first three terms (a quadratic polynomial of $\omega$) in
Eq.(\ref{0+}). Because $\omega_{\max}=10^{6}$Hz$<<T=300$K, factors
$e^{\pm\beta\hbar\omega}\thickapprox1$ do not play a role in the low frequency
regime $\hbar\omega<<k_{B}T$.

To describe the Hall effect, one needs $\psi^{\prime}(t)$ to second order of $J^{field}$: one $J^{field}$ includes
electric field $E_{x}$, and another includes magnetic field $B_{z}$. After
substituting $\psi^{(0)}$, $\psi^{(1)}$ and $\psi^{(2)}$ into the second term
in Eq.(\ref{1e1}), and averaging over various initial conditions,
$\sigma_{yx}$ is determined from $j_{y}$, see Eq.(49) in \cite{4c}. The
primary temperature dependence of $\sigma_{yz}$ is included in the time
integrals which are obtained from integrating out vibrational states. The
3-site processes result to two-fold time integrals\cite{4c} which have
dimension [time]$^{2}$. The time integrals have the same order of magnitude
$s_{AA_{2}A_{1}}^{2}$, where $s_{AA_{2}A_{1}}$ is the characteristic time for
the 3-site processes. Similarly the 4-site processes result to three-fold time
integrals\cite{4c}, the order of magnitude is $s_{AA_{3}A_{2}A_{1}}^{3}$,
where $s_{AA_{3}A_{2}A_{1}}$ is the characteristic time for the 3-site
processes. $s_{AA_{2}A_{1}}$ (3-site processes) and $[\hbar^{-1}%
Js_{AA_{3}A_{2}A_{1}}^{3}]^{1/2}$ (4-site processes) may be explained as the
mean free times in presence of magnetic field. Applying fields $E_{y}$ and
$B_{z}$, $\sigma_{xy}$ is inferred from $j_{x}$. The order $J^{1}$
contributions come from various 4-site processes, their order is $n_{e}%
e^{2}(\hbar^{-1}J_{A_{3}A}s_{AA_{3}A_{2}A_{1}}^{3})\hbar^{-2}$$[B_{z}%
eL_{z}^{A_{2}A_{1}}/m]x_{A_{1}A}v_{A_{3}A_{2}}$, where $L_{z}^{A_{2}A_{1}%
}=\int d\mathbf{r}\phi_{A_{2}}^{\ast}L_{z}\phi_{A_{1}}$ is the matrix element
of the $z$ component of electronic orbital angular momentum. The widely used
3-site process in the literature\cite{fri63,fir,sch65,bot77,mov} is a special
case when $A_{1}=A_{3}$. The order $J^{0}$ contributions comes from 3-site
processes. Their order is $n_{e}e^{2}s_{AA_{2}A_{1}}^{2}\hbar^{-2}(B_{z}%
eL_{z}/m)xv$. It roughly corresponds to the `interference' contribution
(involving two sites)\cite{sch65}.
They are neglected in some other calculations\cite{fri63,hol68}. The ratio of
the order $J^{1}$ terms to $J^{0}$ terms is $(\hbar^{-1}J_{A_{3}A}%
s_{AA_{3}A_{2}A_{1}}^{3})s_{AA_{2}A_{1}}^{-2}\thicksim1$.

\begin{figure}[ptb]
\begin{center}
\includegraphics[width=8cm]{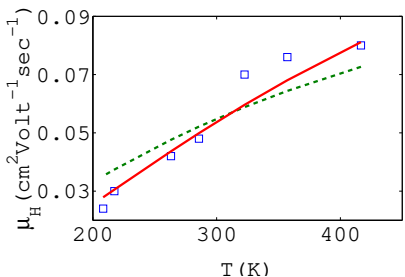}
\end{center}
\caption{Hall mobility vs. temperature: square symbols for n-type
a-Si:H\cite{str}, dashed line from best fit of Friedman-Holstein
formula\cite{fri63}, solid line is from Eq.(\ref{hotem}).}%
\label{Mobhall}%
\end{figure}

Amorphous semiconductors are isotropic, so that one may estimate the Hall
mobility as: $\mu_{H}=B_{z}^{-1}\sigma_{xy}/\sigma_{xx}\thicksim\frac{e}%
{\hbar}xvs_{AA_{2}A_{1}}^{2}t_{A_{1}A}^{-1}$, where $x_{A_{3}A}=i\hbar
v_{A_{3}A}(E_{A}^{0}-E_{A_{3}}^{0})^{-1}$. For a-Si:H\cite{str}, $\xi$ and
$R_{A_{1}A}\thicksim5-10$\AA , $J\thicksim0.02$eV, one has $\mu_{H}%
\thicksim0.1-0.2$cm$^{2}$Volt$^{-1}$sec$^{-1}$. The temperature dependence of
$\mu_{H}$ may be obtained from those of $\sigma_{xy}$ and $\sigma_{xx}$:
\[
\mu_{H}\thicksim\frac{e}{\hbar}xvs_{AA_{2}A_{1}}^{2}t_{A_{1}A}^{-1}%
\]%
\begin{equation}
\exp\{-\frac{E_{A_{1}A}^{a}}{2k_{B}T}-\frac{3}{2}k_{B}T\sum_{\alpha}%
(\hbar\omega_{\alpha})^{-1}(\theta_{\alpha}^{A_{1}}-\theta_{\alpha}^{A}%
)^{2}\}. \label{hotem}%
\end{equation}
At low frequency, the phonon spectral density $\propto\omega^{2}$, so that the
sum in Eq.(\ref{hotem}) converges. Fig.\ref{Mobhall} gives the Hall mobility
vs. temperature for n-type a-Si:H. Both Eq.(\ref{hotem}) and the
Friedman-Holstein result\cite{fri63},
are roughly consistent with experimental data\cite{str}. $E_{A_{1}A}%
^{a}=\lambda_{A_{1}A}(1+\Delta G_{A_{1}A}/\lambda_{A_{1}A})^{2}/4$ is
estimated from typical parameters $\Delta G_{A_{1}A}=0.05$eV and
$\lambda_{A_{1}A}=0.2$eV for a-Si\cite{epjb}. 

We may also demonstrate the undetermined sign of the Hall voltage from Ehrenfest's theorem. The
expected value for the acceleration of a carrier is:
\[
\frac{d}{dt}\int d\tau\psi^{\prime\ast}(t)(\frac{d\mathbf{r}}{dt})\psi
^{\prime}(t)=-\frac{qi\hbar}{2m^{2}}\int d\tau\psi^{\prime\ast}\psi^{\prime
}\nabla\times\mathbf{B}%
\]%
\begin{equation}
+\frac{1}{m}\int d\tau\psi^{\prime\ast}(t)[q\mathbf{E}-\nabla_{\mathbf{r}%
}V(\mathbf{r},\{\mathbf{W}_{\mathbf{n}}\})]\psi^{\prime}(t) \label{ace0}%
\end{equation}%
\[
-\frac{q}{m^{2}}\int d\tau\psi^{\prime\ast}\mathbf{B}\times(-i\hbar
\nabla_{\mathbf{r}}\psi^{\prime})+\int d\tau\psi^{\prime\ast}\frac{q^{2}%
}{m^{2}}(\mathbf{B}\times\mathbf{A)}\psi^{\prime},
\]
where $q$ is the charge of the carrier. If a system has only extended states,
because the mean free time of a carrier is much larger than the effective
interaction time with phonons and defects, the drift velocity is along the
direction of $q\mathbf{E}$ for a nearly free carrier. The direction of the
average magnetic force in an extended state (diagonal element) is the same as
the\ classical one $q\mathbf{E}\times\mathbf{B}$, and the sign of the Hall voltage
is normal. Because the force exerted on a carrier by the external $\mathbf{E}$
field is much weaker than the binding force of the disorder potential, the time
average of $m^{-1}\int\phi_{A_{1}}^{\ast}(-i\hbar\nabla_{\mathbf{r}%
}-q\mathbf{A})\phi_{A_{1}}$ in a localized states $\phi_{A_{1}}$ is zero: no
magnetic force acts on a localized carrier. The mean trapping time that a carrier 
spends in a localized state before making a transition to
other states is\cite{epjb} $\thicksim\hbar J^{-2}(\lambda k_{B}T)^{1/2}%
e^{E_{a}/k_{B}T}$ (high temperature) or $\thicksim\hbar J^{-2}(\Delta G)$ (low
temperature), where $\Delta G$ is the typical energy difference between the
final and the initial states, and $\lambda$ is the reorganization energy for a
transition. The mean transition time needed for a transition event is
$\thicksim mdR/\hbar,$ where $d$ is a typical bond length, $R$ is the distance
between two localized states for LL transition, $R$ is localization length for
localized-extended transition. The mean transition time is much shorter than the mean trapping
time in a typical localized state. Comparing with the transition speed
$\hbar/(md)$, the speed $qEdR/\hbar$ obtained from external electric field
$\mathbf{E}$ during the transition time is negligible. The magnetic force
suffered by a carrier during a transition is along the direction of
$q\mathbf{v}\times\mathbf{B}$, where $\mathbf{v}$ is the transition velocity
of the carrier. $\mathbf{v}$ does not have any relation to the direction of
$\mathbf{E}$. If one applies E field along the x axis and B field along the z
axis, a direction dependent Hall voltage should be detectable along any
direction in the yz plane, not only along the y axis. Checking this prediction
would be a critical test for this work. A recent experiment\cite{cru} shows
that the signs of Hall voltage in several a-Si:H films are not always reverse
to those expected from $q\mathbf{E}\times\mathbf{B}$, the present analysis
seems to agree with this observation.

We show that localized states are responsible for the anomalous sign of the Hall voltage. The
method has the potential to be implemented with current single-particle
\textit{ab initio} simulations, and requires only the eigenvalues and
eigenvectors of the single-electron Hamiltonian, dynamical matrix, and
quantities easily derived from these.

We thank the Army Research Office for support under MURI W91NF-06-2-0026.


\begin{thebibliography}{99}                                                                                               %


\bibitem {ma}A. Miller and E. Abrahams, Phys. Rev. \textbf{120}, 745 (1960).

\bibitem {kik}M. Kikuchi, J. Non-Cryst. Sol. \textbf{59}/\textbf{60}, 25 (1983).



\bibitem {fri71}L. Friedman, J. Non-Cryst. Solids \textbf{6}, 329 (1971).

\bibitem {bot77}H. Bottger and V. Bryksin, Phys. Status Solii b\textbf{80},
569 (1977).

\bibitem {Emin77}D. Emin, \textit{Proc. 7th Int. Conf. on Amorphous and Liquid
Semiconductors}, ed. W. E. Spear (CICL, Edinburgh) 249, (1977).

\bibitem {fri78}L. Friedman and M. Pollak, Phil. Mag. B\textbf{38}, 173 (1978).

\bibitem {mov}B. Movaghar, B. Pohlmann and D. W\"{u}rtz, J. Phys. C: Solid
State Phys. \textbf{14}, 5127 (1981).

\bibitem {str}R. A. Street, \textit{Hydrogenated Amorphous Silicon}, Cambridge
Univresity Press, Cambridge (1991).

\bibitem {motda}N. F. Mott and E. A. Davis, {\it Electronic Processes in
Non-crystalline Materials}, Clarendon Press, Oxford, (1971).

\bibitem {epjb}M.-L. Zhang and D. A. Drabold, arXiv:1004.0404, Eur. Phys. J.
B. DOI: 10.1140/epjb/e2010-00233-0.



\bibitem {hII}T. Holstein, Ann. Phys. \textbf{8}, 343 (1959).

\bibitem {lan62}I. G. Lang and Yu. A. Firsov, Zh. Eksperim i Teor. Fiz.
\textbf{43}, 1843 (1962) [Soviet Phys.-JETP \textbf{16}, 1301 (1963)].

\bibitem {h61}T. Holsetin, Phys. Rev. \textbf{124}, 1329 (1961).

\bibitem {fri63}L. Friedman and T. Holstein, Annals of Physics, \textbf{21},
494 (1963).

\bibitem {fir}Yu. A. Firsov, Fiz. Tverd. Tela \textbf{5}, 2149 (1963) [Soviet
Phys.-Solid State \textbf{5}, 1566 (1964)].

\bibitem {sch65}J. Schnakenberg, Z. Physik. \textbf{185}, 123 (1965).

\bibitem {hol68}T. Holstein and L. Friedman, Phys. Rev. \textbf{165}, 1019 (1968).

\bibitem {urbachletter}Y. Pan, F. Inam, M. Zhang and D. A. Drabold, Phys. Rev.
Lett. \textbf{100} 206403 (2008); P. A. Fedders, D. A. Drabold and S. Nakhmanson,  Phys. Rev. B {\bf 58} 15624 (1998).

\bibitem {attafynn}R. Atta-Fynn, P. Biswas and D. A. Drabold, Phys. Rev. B
\textbf{69} 245204 (2004).

\bibitem {ku57}R. Kubo, J. Phys. Soc. Jpn. 12, 570 (1957).

\bibitem {4c}M.-L. Zhang and D. A. Drabold, arXiv:1008.1067, submitted to Phys. Rev. B.

\bibitem {irv}J. H. Irving and J. G. Kirkwood, J. Chem. Phys. \textbf{18}, 817 (1950).





\bibitem {bcs}J. Bardeen, L.N. Cooper and J. R. Schrieffer, Phys. Rev.
\textbf{108}, 1175 (1957).



\bibitem {kubo}M.-L. Zhang and D. A. Drabold, Phys. Rev. B\textbf{81}, 085210 (2010).

















\bibitem {india}C. J. Mathai, S. Saravanan, M. R. Anantharaman, S.
Venkitachalam and S. Jayalekshmi, J. Phys. D: Appl. Phys. \textbf{35}, 240 (2002).

\bibitem {greece}A. N. Papathanassiou, J. Phys. D: Appl. Phys. \textbf{35},
L88--L89 (2002).

\bibitem {cru}I. Crupi, S. Mirabella, D. D'Angelo, S. Gibilisco, A. Grasso, S.
Di Marco, F. Simone and A. Terrasi, J. Appl. Phys. \textbf{107}, 043503 (2010).


\end{thebibliography}
\end{document}